\documentclass[twocolumn,showpacs,prl,superscriptaddress]{revtex4}

\usepackage{graphicx}
\usepackage{dcolumn}
\usepackage{bm}
\usepackage{color}
\usepackage{amsmath}

\begin{document}

\title{Scaling theory of driven polymer translocation}

\author{T. Ikonen}
\affiliation{Department of Applied Physics and COMP Center of
Excellence, Aalto University School of Science,
P.O. Box 11000,
FI-00076 Aalto, Espoo, Finland}

\author{A. Bhattacharya}
\affiliation{Department of Physics, University of Central Florida, Orlando, Florida 32816-2385, USA}

\author{T. Ala-Nissila}
\affiliation{Department of Applied Physics and COMP Center of
Excellence, Aalto University School of Science,
P.O. Box 11000,
FI-00076 Aalto, Espoo, Finland}
\affiliation{Department of Physics,
Box 1843, Brown University, Providence, Rhode Island 02912-1843}

\author{W. Sung}
\affiliation{Department of Physics, Pohang University of Science and Technology, Pohang 790-784, South Korea}

\date{November 26, 2012}

\begin{abstract}
We present a theoretical argument to derive 
a scaling law between the mean translocation time $\tau$ and the chain length $N$ for driven
polymer translocation. 
This scaling law explicitly takes into account the pore-polymer interactions, which appear as
a correction term to asymptotic scaling and are responsible for the dominant finite size effects in the 
process. 
By eliminating the correction-to-scaling term we introduce a rescaled translocation time 
and show, by employing both the Brownian Dynamics Tension Propagation theory 
[Ikonen {\it et al.}, Phys. Rev. E {\bf 85},  051803 (2012)] and molecular dynamics simulations 
that the rescaled exponent reaches the asymptotic limit in a range of chain lengths
that is easily accessible to simulations and experiments. 
The rescaling procedure can also be used to quantitatively estimate the magnitude of the 
pore-polymer interaction from simulations or experimental data. 
Finally, we also consider the case of driven translocation with hydrodynamic interactions (HIs).
We show that by augmenting the BDTP theory with HIs 
one reaches a good agreement between the theory and previous simulation results found in the literature. 
Our results suggest that the scaling relation between $\tau$ and $N$ is retained even in this case. 
\end{abstract}

\maketitle


{\it Introduction.} 
The transport of a polymer through a nano-sized pore occurs in many biological processes and functions, 
including DNA and RNA translocation through nuclear pores, protein transport across membrane channels 
and virus injection~\cite{albertsbook}. The translocation process is also envisioned to have several 
biotechnological applications in gene therapy, drug delivery and rapid 
DNA sequencing~\cite{meller2003,schadt2010, branton2008,oxfordnanopore}. However, despite significant 
technological advances and considerable experimental~\cite{meller2003, kasi1996, storm2005} and 
theoretical~\cite{chuang2001,kantor2004, sung1996, metzler2003, muthu1999, dubbeldam2007,vocks2008,sakaue2007,sakaue2008,sakaue2010,saito2011, rowghanian2011,luo2008,luo2009,milchev2011,huopaniemi2006,bhatta2009,bhatta2010,bhatta2010b,metzler2010,huopaniemi2007,lehtola2008,lehtola2009, lehtola2010,gauthier2008a,gauthier2008b,saito2012a,saito2012b,luo2007prl,luo2008pre,dehaan2010,grosberg2006,fyta2008} 
studies, until very recently the connection between fundamental theory and numerical simulations 
and experiments has remained elusive (for a review, see, e.g., Ref.~\cite{milchev2011}). 
Recently, we proposed a theoretical model of driven polymer translocation and showed that it 
explains and reproduces all of the results of relevant numerical simulations, 
hence providing a unifying view on the physics of driven polymer translocation~\cite{ikonen2011,ikonen2012}. 
The model couples a Brownian Dynamics (BD) framework for one degree of freedom with the deterministic 
Tension Propagation (TP) theory~\cite{sakaue2007,sakaue2008,sakaue2010,saito2011}. 
The model has so far been used to study driven translocation in the Rouse dynamics picture, 
where it was shown that the pore-polymer interactions induce strong 
finite size effects that persist for extremely long chains up to $N\approx 10^5$~\cite{ikonen2012}.

The detailed formulation of the BDTP theory is rather involved~\cite{ikonen2011,ikonen2012} and except for the
asymptotic large $N$ behavior no explicit analytic solutions to the theory have been found. The main purpose of this 
work is to find a scaling type of solution to the BDTP theory. This is
extremely important in order to obtain an intuitive, physical understanding of the driven translocation process.
We show here that the pore-polymer interactions appear as a {\it correction-to-scaling} term to the asymptotic
behavior of the mean translocation time $\tau$ as a function of the chain length $N$.  
We show how this scaling relation can be used both to understand and quantify the effect of the pore-polymer interactions 
from molecular dynamics (MD) simulations or from experimental data. 
Most importantly, by eliminating the correction-to-scaling contribution
we demonstrate 
that the corresponding 
rescaled exponent reaches its asymptotic limit already for chain lengths $N \lesssim 10^2$ -- 
orders of magnitude faster than the conventional exponent. 
Finally, we discuss the effect of hydrodynamic interactions (HIs) by modifying 
the BDTP model to account for hydrodynamics in an approximate way. 
The effect of the HIs, as calculated from the BDTP model, is to shorten the translocation 
time and the corresponding effective exponent due to the increased (relative) importance 
of the pore-polymer interactions. The predictions of the BDTP model are shown to agree with 
previous simulation studies found in the literature. 
The result suggests that the scaling solution holds even with HIs, 
highlighting the importance of this result.


{\it Theory.} 
Although a details of the BDTP theory are complex, 
some central results can be derived from relatively simple arguments. 
To find the mean translocation time $\tau$ (or the mean translocation velocity $\langle v \rangle$), 
for a chain of length $N$, one needs to consider the force balance between the driving force $f$ 
and the drag force, as suggested, e.g., in Ref.~\cite{storm2005}. 
In principle, the total drag has three contributions: the friction due to the {\it cis} side 
subchain and the solvent, the corresponding friction for the {\it trans} side subchain 
and the friction of the chain portion inside the pore. For driving forces typically used in experiments and simulations, 
the {\it trans} side has an almost negligible contribution to the overall friction~\cite{ikonen2011,ikonen2012}. 
Therefore, the effective friction $\Gamma$ can be approximately written as $\Gamma(t)=\eta_\mathrm{cis}(t) + \eta_p$, 
where $\eta_\mathrm{cis}$ is the friction of the {\it cis} side subchain and $\eta_p$ is the (effective) pore friction. 
If the length of the pore, $l_p$, is small compared to the contour length of the chain, 
$l_p \ll aN$ (with $a$ the segment length), the number of segments occupying the pore and thus
the pore friction $\eta_p$ can be 
regarded as constant during the translocation process.

The friction due to the {\it cis} side subchain, on the other hand, depends explicitly on the 
number of mobile monomers on the {\it cis} side. Due to the nonequilibrium nature 
of the problem, this number depends not only on the chain length $N$ but also on time 
$t$~\cite{lehtola2008,lehtola2009, sakaue2007,sakaue2010,saito2011,ikonen2011,ikonen2012,rowghanian2011,dubbeldam2011,grosberg2006}. 
Solved as a function of time, $\eta_\mathrm{cis}(t)$ shows non-monotonic behavior that 
suggests the division of the translocation process into two stages: the tension propagation stage 
of increasing friction, and the post propagation stage of decreasing friction~\cite{ikonen2011,ikonen2012}. 
However, averaged over the whole process, the time-averaged $\eta_\mathrm{cis}$ is 
approximately given as $\langle \eta_\mathrm{cis} \rangle \sim N^\nu$, 
leading to the asymptotic ($N\rightarrow \infty$) scaling of the mean translocation time as $\tau \sim N^{\nu+1}$~\cite{ikonen2011,ikonen2012,rowghanian2011,dubbeldam2011}.

\begin{figure}
\includegraphics[width=0.8\columnwidth]{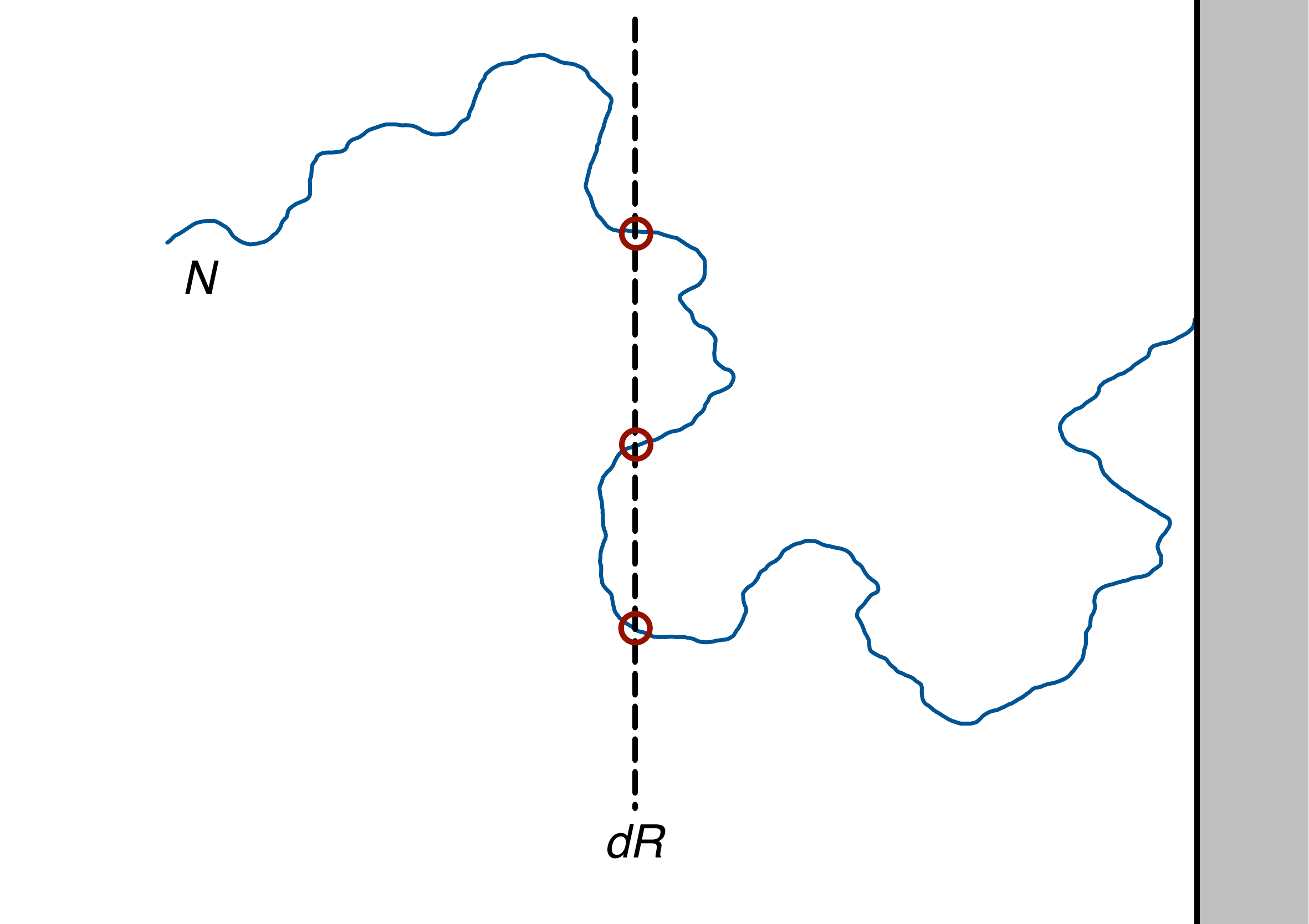}
\caption{(Color online) A schematic illustration of the configuration of a polymer tethered to a wall. 
Intersections of the polymer with an imaginary plane of thickness $dR$ oriented 
parallel to the wall are indicated by the red circles. 
The average number of times a polymer of length $N$ intersects the plane is proportional to $N^{1-\nu}$.}
\label{fig:configuration}
\end{figure}

The full solution of $\eta_\mathrm{cis}(t)$ is rather involved (cf. Refs.~\cite{ikonen2011,ikonen2012}) and
does not give much physical insight. We will now show that 
the result $\langle \eta_\mathrm{cis} \rangle \sim N^\nu$ can be obtained with relatively 
simple argument, based in part on the work of DiMarzio~\cite{dimarzio1979}. 
In equilibrium, the polymer assumes a configuration comprising several loops. 
Before the driving force can be transmitted to any given chain segment, 
the preceding loops need to be straighthened. Only then can the tension 
propagate along the chain backbone. Therefore, at any given time the 
friction $\eta_\mathrm{cis}$ is due to the motion of the unraveling loop closest to the pore, 
while the rest of the chain is essentially immobile. To estimate the average length of the loop, 
we note that for large $N$, the end-to-end distance of the polymer is given by the usual
Flory scaling form $R\sim a N^\nu$. To estimate the number of times the chain intersects a plane of thickness $dR$ parallel to 
the membrane (see Fig.~\ref{fig:configuration}), we note that the number of segments within the plane, $dN$, is proportional to the average line density of monomers, $N/R$, giving $dN\sim N^{1-\nu} dR$.
This is proportional to the number of times the chain intersects the plane and 
also approximately the average number of loops (for large $N$). 
The average length of one loop is thus proportional to $N/N^{1-\nu}=N^\nu$. 
The number of mobile monomers at any given time is therefore proportional to $N^\nu$, 
and the friction due to the drag on the {\it cis} side is $\langle \eta_\mathrm{cis} \rangle \sim \eta N^\nu$, 
where $\eta$ is the solvent friction per chain segment. 
The total effective, time-averaged friction is thus of the 
form $\langle \Gamma \rangle \approx C \eta N^\nu +  \eta_p$, with $C$ an $N$-independent constant.

For a sufficiently strong driving force $f$, one can express the mean 
translocation time in terms of the average translocation velocity $\langle v \rangle$. 
Force balance implies $\langle v \rangle = f/\langle \Gamma \rangle$. Since the driven translocation proceeds via gradual uncoiling of the whole chain, the relevant length scale of the process is the chain's contour length, $aN$ (note that this is in contrast with the assumption of Ref.~\cite{storm2005}, where uniform contraction of the {\it cis} side chain suggests the radius of gyration as the relevant length scale). Thus, 
$\tau = aN/\langle v \rangle = aN\langle \Gamma \rangle /f$. With respect to the chain length, the complete relation is then
\begin{equation}
\tau \approx A(f,\eta)  N^{1+\nu} + B(f,\eta) \tilde{\eta}_p N,\label{eq:tau}
\end{equation}
where $A$ and $B$ are independent of $N$, 
and $\tilde{\eta}_p \equiv \eta_p / \eta$ is the dimensionless pore friction. 
This scaling form is one of the main results in this paper. It shows that the
influence of the pore friction appears as a {\it correction-to-scaling} to the
asymptotic value of
the translocation exponent $\alpha$, defined via $\tau\sim N^\alpha$, 
which approaches $\alpha_\infty = 1+\nu$ from below. 
In addition, Eq.~(\ref{eq:tau}) suggests that if the dimensionless pore friction increases 
(e.g., by either decreasing the solvent friction $\eta$ or by reducing the pore radius), 
the exponent $\alpha$ becomes smaller, especially for relatively short chains, in agreement
with theory~\cite{ikonen2012} and MD simulations~\cite{bhatta2010b, huopaniemi2007,lehtola2010,dehaan2010}.

\begin{figure}
\includegraphics[width=0.95\columnwidth]{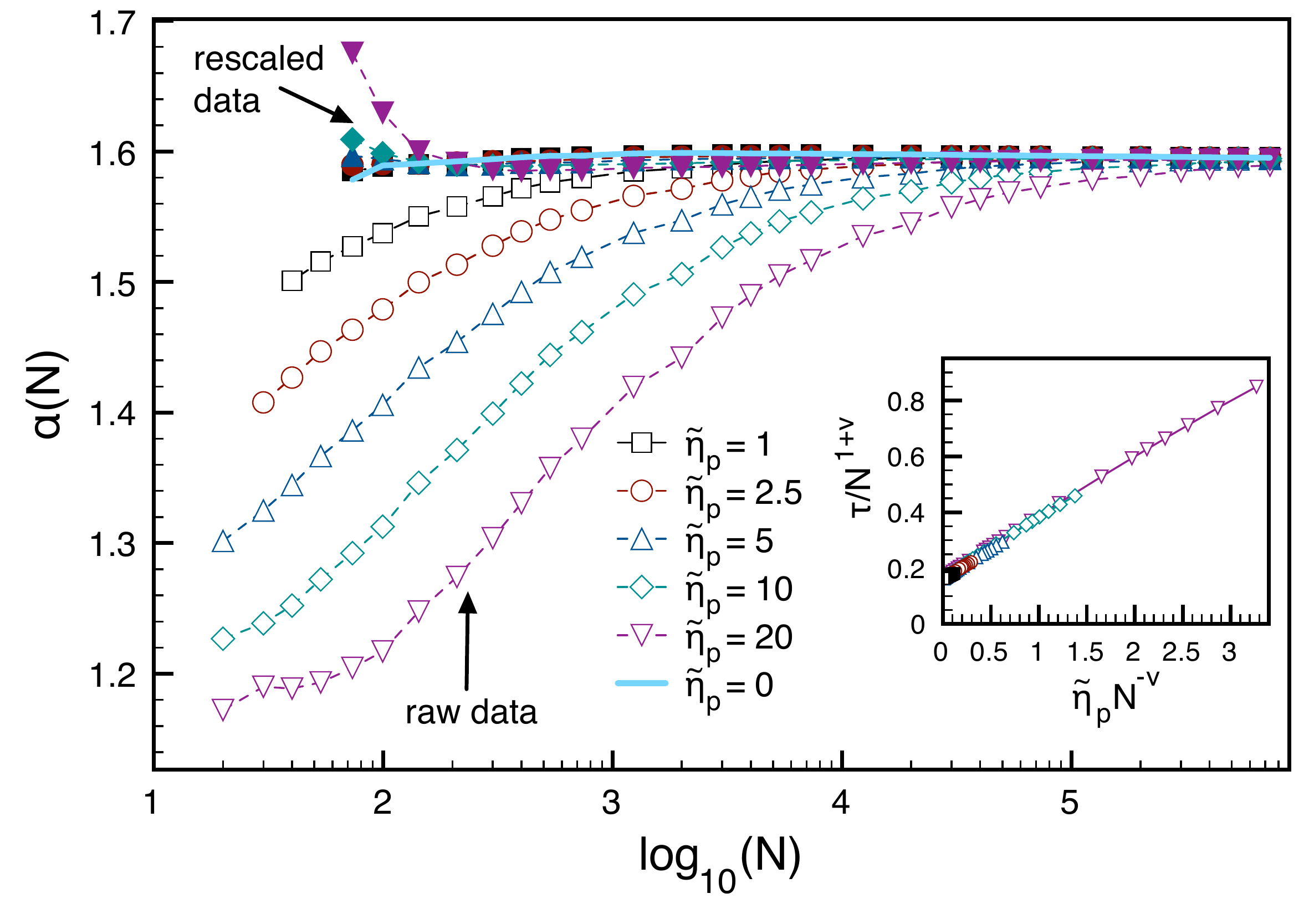}
\caption{(Color online) Open symbols, main plot: the effective exponent  $\alpha(N)$ as a function of 
the chain length $N$ for the BDTP model solved for different ratios $\tilde{\eta}_p$ of pore and 
solvent friction. Solid symbols: the same data rescaled to exclude the effect of pore friction, 
showing collapse of the data to the asymptotic exponent $1+\nu$.  
Inset: the data collapse for different $\tilde{\eta}_p$ on a ($\tau/N^{1+\nu}, \tilde{\eta}_pN^{-\nu}$) 
plot used to extract the finite size effect of the pore friction (see text for details). 
Numerical errors are of the order of the symbol sizes, or smaller.}
\label{fig:alpha_BDTP}
\end{figure}

{\em Results.} 
To clarify the influence of the pore friction on the translocation time, we have solved $\tau$ 
time as a function of chain length for different pore frictions using the BDTP model~\cite{ikonen2011,ikonen2012}. 
Using $\tau(N) \sim N^\alpha$, one can extract the effective scaling 
exponent $\alpha(N)$ as $\alpha(N) = \frac{d\ln \tau}{d\ln N}$. 
Because of the linear sub-leading term in Eq.~(\ref{eq:tau}), 
$\alpha$ has a weak dependence on the chain length, and approaches $1+\nu$ extremely slowly
with typical values of the parameters. 
Naturally, for larger $\tilde{\eta}_p$, the dependence on $N$ is more pronounced, 
as shown in Fig.~\ref{fig:alpha_BDTP}. 
The other parameter values used in solving the BDTP model were 
$f=5.0$ (driving force), $k_BT=1.2$ (temperature), and $\nu=0.588$.

To quantitatively show that the deviation from the asymptotic limit is caused by the pore friction, 
we subtract the correction-to-scaling term and define a rescaled translocation time 
${\tau}^\dagger$ as ${\tau}^\dagger={\tau}-B\tilde{\eta}_p N$ and the 
corresponding rescaled exponent as $\alpha^\dagger(N)\equiv\frac{d\ln {\tau}^\dagger}{d\ln N}$.  According to Eq.~(\ref{eq:tau}), the rescaled translocation time can be expressed as
\begin{equation}
\tau^\dagger = \left( {\tau}-B\tilde{\eta}_p N \right) \sim N^{\alpha^\dagger},
\end{equation}
with the exponent $\alpha^\dagger$  independent of $N$ at $\alpha^\dagger \approx 1  + \nu$. 
In practice, to obtain the rescaled ${\tau}^\dagger$, one has to find the numerical prefactor $B$ 
by means of finite size scaling. In the inset of Fig.~\ref{fig:alpha_BDTP} this is done by plotting the 
translocation time data in the form $\tau/N^{1+\nu} = A + B\tilde{\eta}_pN^{-\nu}$. 
The coefficients $A$ and $B$ can then be obtained from a simple linear least squares fit.

The rescaled exponent $\alpha^\dagger$ as obtained from the BDTP model is shown in 
Fig.~\ref{fig:alpha_BDTP} as solid symbols. All the curves corresponding to different values of 
$\tilde{\eta}_p$ {\it collapse onto a single master curve} 
within the numerical accuracy 
around $\alpha^\dagger = 1+\nu$, also coinciding with the ideal $\tilde{\eta}_p = 0$ solution. 
For very short chains, the collapse is not perfect because of secondary finite chain 
length effects that come into play, 
and also because as a continuum level description 
the BDTP model may not accurately describe very short chains.
 
\begin{figure}
\includegraphics[width=0.95\columnwidth]{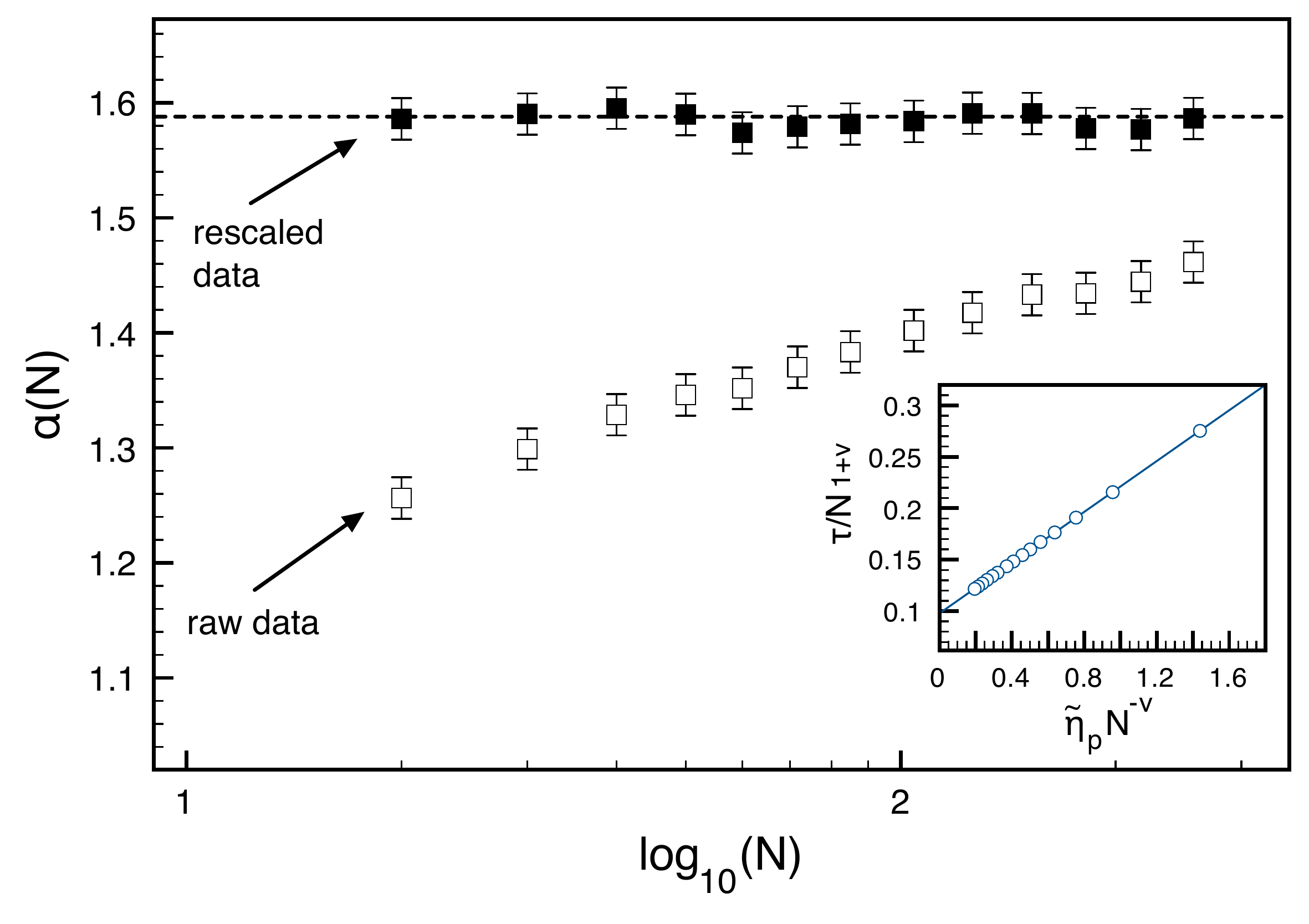}
\caption{(Color online) Open symbols, main plot: the effective exponent  $\alpha(N)$ as a 
function of chain length $N$ as obtained from MD simulations 
(for details, see text). Solid symbols: the same data rescaled as in Fig.~\ref{fig:alpha_BDTP}, 
showing collapse to the value $1+\nu$ (dashed line).  
Inset: Finite size scaling plot used to extract the contribution of the pore friction.}
\label{fig:alpha_MD}
\end{figure}

We have confirmed that the rescaling scheme works for raw data obtained from MD simulations, too, by 
performing Langevin thermostatted MD simulations using the Kremer-Grest bead-spring model~\cite{kremer-grest} 
with typical parameters found in the literature 
(for a more detailed description, see, e.g., Refs.~\cite{ikonen2011,ikonen2012}). 
We used the parameters $f=5.0$, $k_BT=1.2$, $\eta=0.7$, 
and pore diameter $d_p=2.0$ and length $l_p=1.0$ (both in units of  the segment length $a$), 
which correspond to the dimensionless pore friction of about $\tilde{\eta}_p\approx 5.6$~\cite{ikonen2012}. 
The chain lengths in the simulations were $10\leq N \leq 300$, with $\tau$ averaged over at least $5000$ successful events for each $N$.
 
Plotting the effective exponent $\alpha(N)$ reveals the same 
dependence on $N$ as predicted by the theory. Furthermore, by performing finite size scaling 
similar to the BDTP case, we see that the rescaled exponent $\alpha^\dagger$ reaches the asymptotic 
value almost immediately; within the statistical error, the value is $\alpha^\dagger=1+\nu$ already for $N=20$. 
This shows that Eq.~(\ref{eq:tau}) is valid for remarkably short chains and that the dominant finite size effect 
in the driven translocation process is the frictional interaction between the pore and the polymer.
 
Finally, we would like to point out another important feature of the scaling solution.
Used in reverse, the finite size scaling procedure can be used to estimate $\tilde{\eta}_p$ 
if its value is unknown for the pore geometry in question. 
By varying $\tilde{\eta}_p$ so that the rescaled exponent becomes independent of $N$ 
at $\alpha^\dagger \approx 1 + \nu$, the pore friction can be measured for any geometry 
that satisfies $l_p \ll aN$. While measuring the pore friction directly from the monomer 
waiting time distribution as outlined in Refs.~\cite{ikonen2011,ikonen2012} is more accurate, 
the reverse scaling procedure may be used even if the waiting time distribution is not available, 
as is typically the case in experiments.

Because the driven polymer translocation problem is inherently a dynamical, non-equilibrium process, it is expected
that hydrodynamic interactions (HIs) from solvent should play a role.
In the simplest approximation, the HIs can be included by considering the Zimm type of friction instead of 
the Rouse friction as in Langevin dynamics simulations. 
This means writing down the force balance condition such that the drag force (and the friction $\eta_\mathrm{cis}$) 
is proportional to the linear length of the mobile subchain on the {\it cis} side, 
instead of being proportional to the number of mobile monomers. 
The rationale is that for intermediate forces, when the {\it cis} side subchain 
adopts a shape reminiscent of a trumpet (or a stem-and-flower), 
the solvent inside the trumpet is also set in motion and therefore the 
monomers inside the trumpet do not fully contribute to the drag force. 
Therefore, the overall drag force from the {\it cis} side subchain should be 
slightly smaller for the Zimm case. However, since the maximum linear length of the mobile part of the {\it cis} side
subchain is still given by the end-to-end distance $R\sim a N^\nu$, 
we expect Eq.~(\ref{eq:tau}) to hold even in this case, 
although with a smaller prefactor $A$ in the leading term.

To test this simple argument, we have implemented the Zimm friction of the {\it cis} 
side subchain into the BDTP model. To this end, 
one merely modifies the force balance equation [e.g., Eq. (A1) in Ref.~\cite{ikonen2011}] so 
that the the drag force is given by the size of the tension blob ($\xi$) 
instead of the number of monomers within the blob ($\xi^{1/\nu}$). 
This amounts to replacing the force balance equation of Rouse type as 
$f_\mathrm{drag}^\mathrm{Rouse}(x) =
\frac{1}{a} \int_{-X}^{x} \eta v(x')[\xi(x')/a]^{1/\nu}dx' \rightarrow f_\mathrm{drag}^\mathrm{Zimm} (x)=
\frac{1}{a} \int_{-X}^{x} \eta v(x')dx'$, where $v(x)$ is the instantaneous velocity of the monomers at 
position $x$, and $x=-X$ is the location of the last mobile monomer. One then carries out the same 
numerical implementation as described in Refs.~\cite{ikonen2011,ikonen2012} using the modified force 
balance equation. The difference between the Zimm and Rouse frictions in the context of the tension 
propagation formalism is also further discussed in 
Refs.~\cite{sakaue2007,sakaue2010,saito2011,dubbeldam2011,rowghanian2011}.

\begin{figure}
\includegraphics[width=0.95\columnwidth]{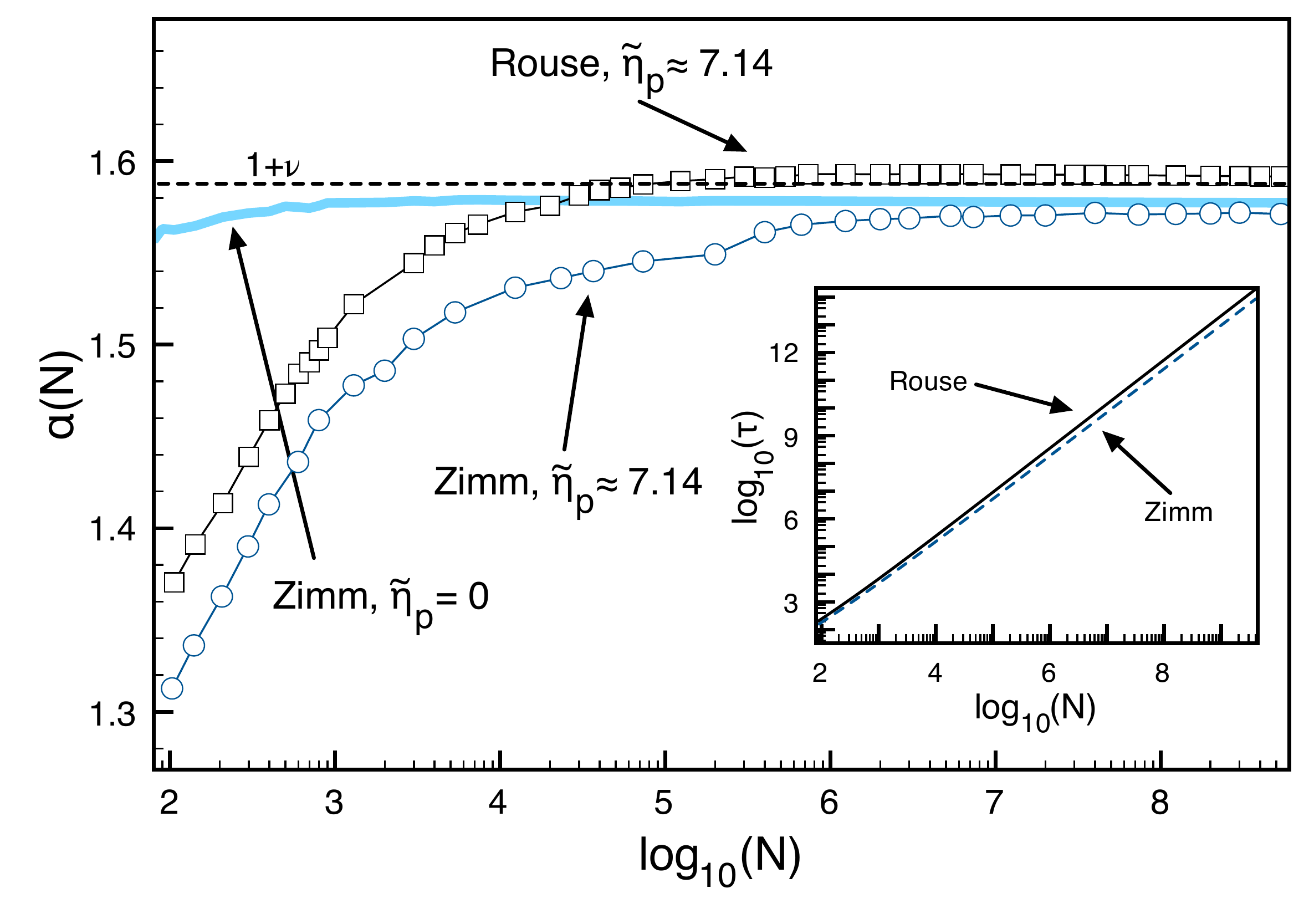}
\caption{(Color online) Main plot: The effective exponent  $\alpha(N)$ as a function of chain length 
$N$ for the BDTP model with Rouse friction (black squares) and Zimm friction (blue circles) 
for $\tilde{\eta}_p\approx 7.14$, and for Zimm friction and $\tilde{\eta}_p = 0$ (light blue solid line). 
Inset: The mean translocation time $\tau$ as a function of chain length $N$ for the Rouse 
(solid black line) and Zimm friction (blue dashed line) for $\tilde{\eta}_p\approx 7.14$. 
Numerical errors are of the order of the symbol sizes, or smaller.}
\label{fig:alpha_hydro}
\end{figure}

The comparison between the Rouse and Zimm dynamics is shown in Fig.~\ref{fig:alpha_hydro}. 
In both cases the BDTP model was solved with 
$f=5.0$, $k_BT=1.2$, $\eta=0.7$, and $\tilde{\eta}_p\approx 7.14$ for up to $N=5\cdot 10^{9}$. 
For the Rouse friction, the effective exponent $\alpha(N)$ is consistently larger than for the Zimm friction, 
and eventually approaches $1+\nu$ for large $N$. For the Zimm friction, the approach to the symptomatic 
limit is considerably slower. In fact, even for $N\approx 10^{9}$, the numerical value ($\alpha\approx 1.57$) 
still increases with $N$. To illustrate the importance of the pore friction term to the slow convergence, 
we have solved the model with zero pore friction ($\tilde{\eta}_p=0$), showing considerably faster 
convergence. Altogether, the numerical results in the large $N$ limit seem to be in agreement 
with $1+\nu$, which has also been predicted to be the asymptotic value using 
analytical approximations to the tension propagation theory~\cite{dubbeldam2011,rowghanian2011}.

In the short chain regime, we obtain $\alpha(N=100) \approx 1.37\pm 0.01$ and $\alpha(N=100) \approx 1.31\pm 0.01$ 
for the Rouse and Zimm cases, respectively. Although a detailed comparison with experiments or 
hydrodynamical simulations is difficult due to the lack of knowledge on 
the pore friction ($\tilde{\eta}_p$), the numerical value of $1.31$ seems to be in good 
agreement with the experimental ($\alpha\approx 1.27\pm 0.03$, Ref.~\cite{storm2005}) 
and lattice-Boltzmann simulation results ($\alpha\approx 1.28\pm 0.01$, Ref.~\cite{fyta2008}). 
In particular, the difference in the exponents measured with and without HIs ($0.06\pm0.02$) 
matches the difference reported in Ref.~\cite{fyta2008} ($0.08\pm0.04$). 
In addition, in agreement with computer simulations~\cite{fyta2008,lehtola2009}, 
the overall translocation time is reduced by the addition of the HIs, 
as shown in the inset of Fig.~\ref{fig:alpha_hydro}.

{\it Conclusions.} 
In this work, we have proposed using theoretical scaling arguments that the mean translocation 
time of a polymer driven through a nanopore can be written in the form 
$\tau \approx A(f,\eta)N^{1+\nu} + B(f,\eta)\tilde{\eta}_pN$. The first term is derived from 
the out-of-equilibrium dynamics of the {\it cis} side subchain and dominates for large $N$, 
while the second term stems from the interactions between the polymer and the pore and 
remains significant for the typical chain lengths in both experiments and computer simulations. 
This unified scaling form is an important physical result and powerful tool for analysis of driven translocation.
By eliminating the correction-to-scaling term 
one can isolate and quantify the effect of pore friction by means of finite size scaling. 
We have demonstrated by using both a theoretical model of translocation dynamics and 
molecular dynamics simulations that the rescaled exponent reaches the asymptotic limit 
already for extremely short chains ($N<100$), whereas the conventionally defined 
exponent does not. In addition, we argue that in the presence of hydrodynamic interactions, 
the translocation time becomes shorter but can still be expressed as a sum of the two terms. 
We present results from the theoretical model proposed in Refs.~\cite{ikonen2011,ikonen2012} 
with hydrodynamic interactions, obtaining quantitative agreement in the scaling exponent $\alpha$ 
with both theoretical and experimental results reported in the literature.


This work has been supported in part by the Academy of Finland through its COMP Center of Excellence
Grant no. 251748. TI acknowledges the financial support of the Vilho, Yrj{\"o} and Kalle V{\"a}is{\"a}l{\"a} Foundation. 
AB has been partially supported by the NSF-CHEM grant \#0809821. 
The authors also wish to thank CSC, the Finnish IT center for science, for allocation of computer resources.

\end{document}